# Mechanism of Electric Power Generation from Ionic Droplet Motion on Polymer Supported Graphene

*Shanshan Yang[1,†], Yudan Su[1,†], Ying Xu[1], Qiong Wu[1], Yuanbo Zhang[1,2], Markus B. Raschke[3], Mengxin Ren[4], Yan Chen[5], Jianlu Wang[5], Wanlin Guo[6], Y. Ron Shen[1,7] and Chuanshan Tian[1,2,\*]*

[1] *Department of Physics, State Key Laboratory of Surface Physics and Key Laboratory of Micro- and Nano-Photonic Structures (MOE), Fudan University, Shanghai, 200433, China*

[2] *Collaborative Innovation Center of Advanced Microstructures, Nanjing, 210093, China*

[3] *Department of Physics, Department of Chemistry, and JILA, University of Colorado, Boulder, CO 80309, United States*

[4] *School of Physics and TEDA Applied Physics Institute, Nankai University, Tianjin, 300071, China*

[5] *National Laboratory for Infrared Physics, Shanghai Institute of Technical Physics, Chinese Academy of Science, Shanghai, 200083, China*

[6] *Institute of Nanoscience, Nanjing University of Aeronautics and Astronautics, Nanjing 210016, China.*

[7] *Department of Physics, University of California, Berkeley, CA 94720, United States*

\* Corresponding author: cstian@fudan.edu.cn
† These authors contributed equally to this work.

**Abstract**

Graphene-based electric power generation that converts mechanical energy of flow of ionic droplets over the device surface into electricity has emerged as promising candidate for a blue-energy network. Yet the lack of a microscopic understanding of the underlying mechanism has prevented ability to optimize and control the performance of such devices. This requires information on interfacial structure and charging behavior at the molecular level. Here, we use sum-frequency vibrational spectroscopy (SFVS) to probe the interfaces of devices composed of aqueous solution, graphene and supporting polymer substrate. We discover that the surface dipole layer of the polymer is responsible for ion attraction toward and adsorption at the graphene surface that leads to electricity generation in graphene. Graphene itself does not attract ions and only acts as a conducting sheet for the induced carrier transport. Replacing the polymer by an organic ferroelectric substrate could enhance the efficiency and allow switching of the electricity generation. Our microscopic understanding of the electricity generation process paves the way for the rational design of scalable and more efficient droplet-motion-based energy transducer devices.

**Introduction**

The graphene-electrolyte interface has been demonstrated to exhibit promising attributes as platform for a range of energy devices, such as solar cells[1], super-capacitors[2], and lithium-ion batteries[3]. Notably, a prototype of a novel graphene-based electric generator has recently been invented using a graphene-liquid interface to convert mechanical energy of moving ionic droplets to electric energy; it offers an attractive new scheme for scalable electric power generation.[4-14] In such a device, droplets or waves of an ionic solution moving across graphene supported by an appropriate substrate generate a current in graphene along or opposite to the flow direction. More recently, such effect can also be observed at the aqueous interface with a polymer coated insulator-semiconductor structure.[15] Macroscopically, the governing mechanism of such electrokinetic phenomenon can be explained by a drawing potential model.[5] It suggests that selective ions from solution can adsorb at the solid/solution interface and form a pseudo capacitor with the solid. As an ionic droplet moves along a graphene surface, ions that tend to adsorb on the interface are attracted towards the advancing front (charging of the pseudo capacitor) or repelled from the receding edge (discharging of the pseudo capacitor). Concurrently, oppositely charged carriers in graphene are attracted to the advancing and receding edge, resulting in a current flow in graphene. Thus how effectively ions can be attracted to the interface determines the efficiency of electricity generation. On the microscopic scale, however, there are still arguments on the underlying mechanism that attracts ions to the solution/graphene interface.[5,8,9,16,17] This current lack of microscopic understanding hinders our ability to optimize and control the performance of such graphene-based energy transducers.

Most work to date has been focused on polymer-supported graphene devices. No consensus has been reached on the fundamental question of how the interface attracts the electrolyte ions to graphene surface. Several mechanisms have been proposed. Based on density functional theory (DFT) calculations, some suggest that cations ($Na^+$) from electrolytic solution would preferentially adsorb on graphene.[5,16,17] Others believe that a polymer substrate could be pre-charged by various means during sample preparation, for example, through friction before transferring of graphene on it;[8] the pre-charged polymer could attract ions from solution to graphene since field screening by the monolayer graphene is weak.[18] For future development of such devices, it is imperative to pinpoint the mechanism of ion-substrate interaction and learn about the relevant parameters. This requires a molecular-level interfacial study of the device, and sum-frequency vibrational spectroscopy (SFVS) is known to be a unique analytical tool for liquid/solid interfaces.[19]

We report here a SFVS study on polymer-supported graphene-based electricity generation devices. From SF vibrational spectra of polymer surfaces, solution/polymer interfaces, and solution/graphene/polymer interfaces, we find conclusively the following: Ions from the solution are not attracted by graphene; they are attracted to graphene/polymer, or polymer without graphene, by surface dipoles of the polymer; the monolayer graphene appears as a weak screening layer for the dipole field and serves as a passive conductive path for the generated current; and the interaction between ions and the surface dipole layer is of short range. Our results provide a more comprehensive picture of electricity generation by the electrolytic solution/graphene/polymer devices that would help in future design of such devices for better efficiency and switchable operation.

**Electricity generation by moving droplets on graphene: dependence on supporting polymer.**

Figure 1a describes the device and the experimental arrangement in our study. To avoid complication caused by possible contamination of graphene,[20] we developed a new method, different from the traditional ones,[21] to lay graphene on polymer. The procedure is described in Method and Fig. S1 of the Supporting Information (SI). Poly-ethylene terephthalate (PET), poly-methyl methacrylate (PMMA), and poly-vinylidene fluoride (PVDF) were studied in our experiment.

To test electricity generation of the device, we adopted the falling droplet scheme (Fig. 1b).[8] Successive drops of 600mM NaCl solution, 5mm in diameter, were let to roll down the graphene surface of the device, which was tilted 60° with respect to the horizontal. Their initial speed on graphene was controlled by the height H (Fig. 1b), set at 10 cm. Voltage pulses generated across the graphene by the moving drops were recorded by an oscilloscope. As seen in Fig 1c, positive voltage spikes (corresponding to positive ions attracted to the water/graphene interface) generated by sequential drops on graphene/PET were readily observed, but were not detectable on graphene/PMMA. This result clearly suggests that it is the polymer substrate, rather than graphene, that attracts ions to the water/graphene interface, and PET attracts $Na^+$ much more strongly than PMMA.

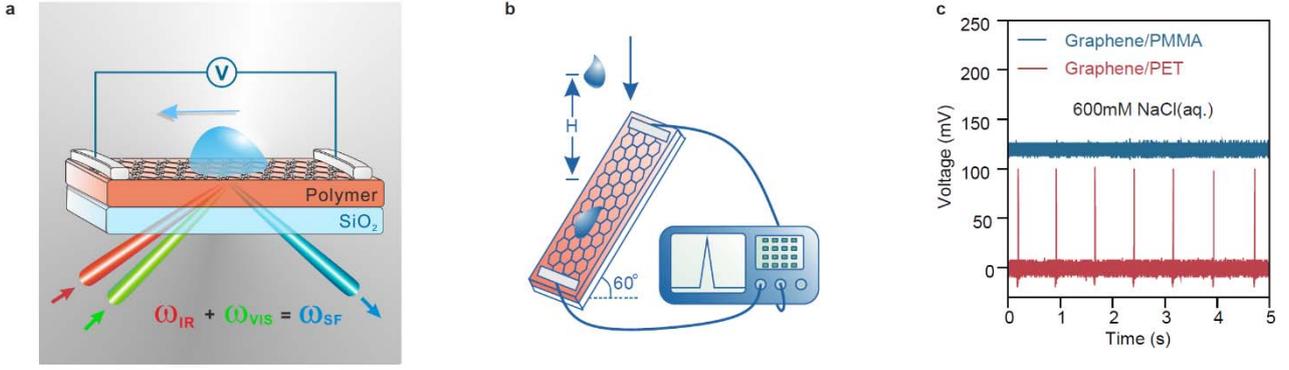

**Figure 1 Electric power generation in polymer supported graphene device.** (a) Experimental arrangement for electricity generation and SFVS measurements on a device composed of a graphene/polymer film supported by a SiO$_2$ plate. (b) Cartoon describing the measurement of voltage generation by a falling aqueous drop rolling on the graphene/polymer device. (c) Oscilloscope traces showing presence and absence of voltage spikes generated from graphene/PET (red) and graphene/PMMA (blue, offset for clarity), respectively. The drops were from a 600mM NaCl aqueous solution.

### SFVS probing of surface field induced by adsorbed ions at the interfaces

Sodium ion adsorption at the interface of water in contact with graphene/polymer can be monitored by SFVS. It is known that a surface field at a water interface can reorient water molecules in the interfacial layer, and result in a spectral change that can be detected by SFVS.[22-28] The spectroscopic technique measures the OH stretch spectrum of the effective surface nonlinear susceptibility $\chi^{(2)}_{S,eff}$ of the interfacial water, which has the expression[29,30]

$$\chi^{(2)}_{S,eff}(\omega) = \chi^{(2)}_S + \int_{0^+}^{\infty} \chi^{(3)}_B E_{DC}(z) e^{i\Delta k_z z} dz \qquad (1)$$

where $\chi^{(2)}_S$ denotes contribution from a few monolayers of water molecules right at the interface, the integral describes the contribution from field-induced polarization of water molecules in the diffuse layer, $E_{DC}(z)$ is the distance dependent surface field along the surface normal, $\chi^{(3)}_B$ denotes the third-order nonlinear susceptibility of bulk water, and $\Delta k_z$ is the phase mismatch of the SFVS process. The change of $\chi^{(2)}_{S,eff}(\omega)$, in particular $\text{Im}\,\chi^{(2)}_{S,eff}(\omega)$, directly reflects the change of $E_{DC}(z)$ in both magnitude and direction.

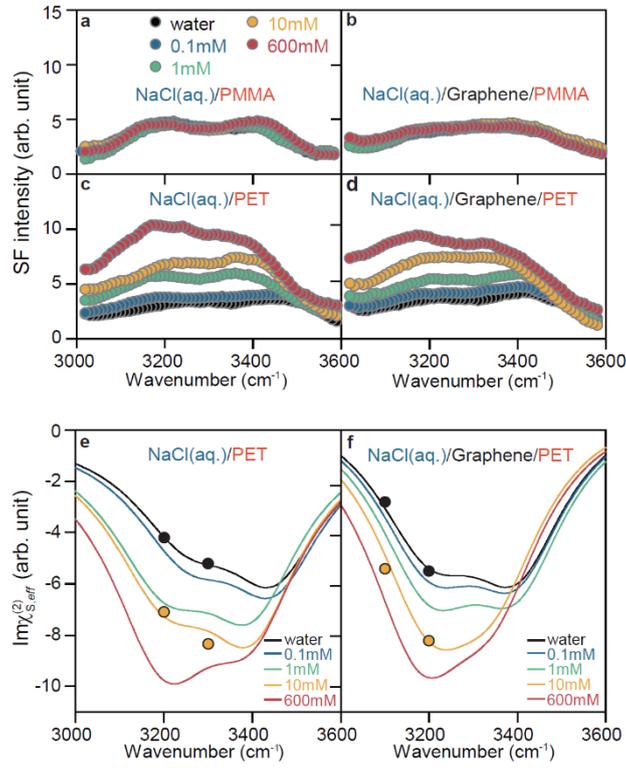

**Figure 2. Bonded OH stretch SFVS spectra of interfacial water.** (a) and (b) SF intensity spectra from the interfaces of NaCl(aq.)/PMMA and NaCl(aq.)/graphene/PMMA, respectively, for a set of different NaCl concentrations in the solution. The nearly identical spectra for all concentrations indicate that both interfaces are essentially neutral, with no ion adsorption. In contrast, the SF intensity spectra from the interfaces of (c) NaCl(aq.)/PET and (d) NaCl(aq.)/graphene/PET increase with NaCl concentration, indicating an increasing density of ions at the interfaces. A small field screening effect of graphene is seen from the lower spectral intensity in (d) in comparison to (c). The $\text{Im}\chi^{(2)}_{S,eff}$ spectra for (e) NaCl(aq.)/PET and (f) NaCl(aq.)/graphene/PET interfaces, deduced from the corresponding SF intensity spectra in (c) and (d) and discrete phase points in (e) and (f) measured directly by phase-sensitive SFVS, become increasingly negative with increasing NaCl concentration, indicating an increasing amount of $Na^+$ ions adsorbed at the interface.

We conducted SFVS measurements on solution/polymer and solution/graphene/polymer interfaces. The existence of a surface field created by PET, but not by PMMA, can be seen from the spectral variation with increasing salt concentration in water. Ions in water can move toward the water interface in response to the surface field and modify the field. Figure 2a-d displays the spectra for the four aforementioned interfaces with different salt concentrations in water ranging from 0 to 600mM. Within measurement error, the spectra of water/PMMA and water/graphene/PMMA are independent of the salt concentration (Fig. 2a and 2b). Obviously, ions do not come to the interface to alter the interfacial water structure. This is a clear manifestation that neither graphene nor PMMA attracts ions. The case of PET is different. The spectrum changes significantly with salt concentration (Fig. 2c and 2d); Na$^+$ ions must have come to the interface to perturb the water structure. The spectrum of the water/graphene/PET interface at each salt concentration is slightly lower than that of the water/PET interface, indicating a weak screening effect of graphene.

The positive ion attraction to graphene was speculated to be caused by pre-existing negative surface charges on the substrate before graphene was transferred onto it.[8] However, knowing that Na$^+$ ions do not spontaneously adsorb on graphene, a negative surface charge layer on the polymer that attracts Na$^+$ would have set up an electric double layer (EDL) with a negative surface field in the adjoining water and make Im $\chi^{(2)}_{S,eff}$ more positive. In contrast, a close look at the Im $\chi^{(2)}_{S,eff}$ spectra of the water/PET and water/graphene/PET interfaces for different salt concentration in water reveals that they become more negative at higher salt concentration (Fig. 2e and 2f).[31] This means that the surface field is positive and increases with salt concentration, orienting the interfacial water molecules with O→H more toward the bulk water, opposite to what is expected from a negatively pre-charged surface. This observation can only be explained by adsorption of Na$^+$ ions at the interface creating a positive surface field.

**Origin of ion adsorption at the interface**

With pre-existence of negative surface charges on PET out of the question, what is then the origin of Na$^+$ ion adsorption at the interface? The surface of a polymer can often be polar with certain molecular group polar-oriented at the surface. If the group has a strong dipole, the polymer should possess a strong surface dipole layer that can attract ions. SFVS can be used to learn about the polar surface structure of a polymer buried under graphene. For PMMA with a chemical formula [CH$_2$=C(CH$_3$)CO$_2$CH$_3$]$_n$, it has been found that the side chains -CO$_2$CH$_3$ dominate on the surface with CH$_3$ projected out at 30° from the surface normal.[32] However, CH$_3$ has a very weak dipole and accordingly, PMMA has a very weak surface dipole layer. In contrast, PET with [C$_8$H$_8$(CO$_2$)$_2$]$_n$ has the carbonyl groups (C=O) normally protruding out of the surface.[33] Because C=O has a very strong dipole, PET has a strong dipole layer. Figure 3a shows the C=O stretch spectra of PMMA and PET surfaces covered by graphene, obtained by SFVS. The very prominent C=O peak at ~1725 cm$^{-1}$ for PET illustrates the strong polar ordering of C=O on the PET surface, while the undetectable C=O peak for PMMA indicates little polar ordering of C=O along the surface normal. The surface dipole layer of C=O with O pointing out on PET must be responsible for attraction of Na$^+$ to the water/graphene interface.

Now the question is how a surface dipole layer can attract Na$^+$ from solution to adsorb on graphene/PET. It is well known, from the continuum theory of electrostatics, that an infinite continuum of surface dipole layer has no field outside the layer. In reality, however, surface dipoles associated with molecular groups reside at discrete positions on the surface. As pointed out by Cherepanov *et al.*, while such local dipoles are of very short range, they create a local potential well on top of each dipole that can trap ions.[34] For illustration, we plot, in Fig. 3b, the calculated potential distribution created by a square array of surface dipoles with a surface density equal to that of C=O on PET (see SI for details). Positive ions adsorbed in the potential traps, now play the dominant role in setting up the positive surface field and the electric double layer in water at the interface. Because the surface dipole field is of short range (see Fig. 3b), the screening effect of graphene can be appreciable. Accordingly, ion adsorption decreases significantly with increase of the graphene layer thickness. This is illustrated in Fig. 3c, where it is seen that the increase of the SF signal with ion concentration is much less for the 3-5 layer graphene/PET than for the monolayer graphene/PET.

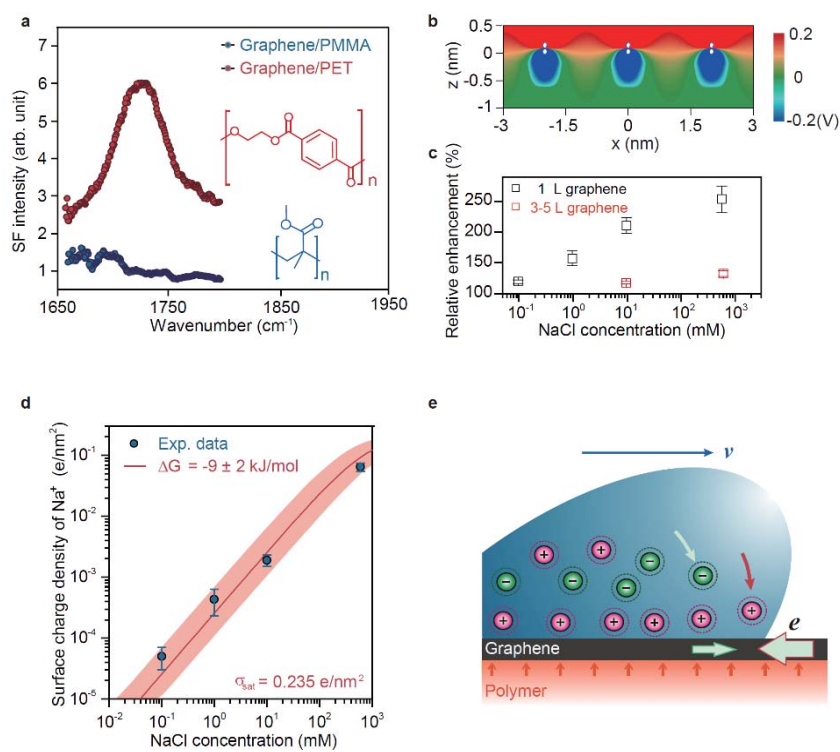

**Figure 3 Origin of ion adsorption at water/graphene/PET interface.** (a) SF vibrational spectra showing the absence and presence of the C=O stretching mode at the graphene/PMMA and graphene/PET interfaces, respectively, denoting strong C=O polar ordering on the latter interface. (b) Calculated azimuthally isotropic electric potential in the x-y plane generated by a square array of C=O dipoles with C→O pointing normally toward the solution. It appears as highly localized potential

wells that can trap positive ions. (c) Relative increase of SF signals for monolayer and 3-5 layers of graphene/PET devices versus NaCl concentration. The SF signals are normalized against that of pure water. (d) Surface Na$^+$ density versus bulk NaCl concentration deduced from the Im $\chi_{S,eff}^{(2)}$ spectra using the Gouy-Chapman model. Fitting of the data points by a simple Langmuir isotherm, with the surface dipole density taken as the saturated ion density $\sigma_{sat}$, allows the deduction of the adsorption free energy $\Delta G$. (e) Cartoon illustrates the motion of ions in solution and electrons in graphene toward the front edge of the droplet in response to an ionic droplet moving forward with velocity v..

The adsorption energy of ions in the surface dipole traps is expected to be small for the C=O type surface dipoles. We can have an estimate on the adsorption energy from the observed spectral variation of interfacial water with different salt concentrations. As seen in Eq.(1), in the limit of low surface density of adsorbed ions (Na$^+$), only the second term on the right depends on the surface field $E_{DC}(z)$, which is generated by the adsorbed ions. With the help of the Gouy-Chapman model, we can relate $E_{DC}(z)$ to the surface ion density $\sigma$, and use it as an adjustable parameter to calculate $\Delta \operatorname{Im} \chi_{S,eff}^{(2)} = \operatorname{Im}[\chi_{S,eff}^{(2)} - \chi_S^{(2)}] = \operatorname{Im} \int_{0^+}^{\infty} \chi_B^{(3)} E_{DC}(z) e^{i\Delta k_z z} dz$ and fit the measured $\Delta \operatorname{Im}\chi_{S,eff}^{(2)}$ with known $\chi_B^{(3)}$ (SI). We can thus find $\sigma$ for the different bulk ion concentrations specified in Fig. 2. The data points are plotted in Fig. 3d and fitted by a simple Langmuir adsorption isotherm with the assumption that the saturated surface ion density, $\sigma_{sat}$, is equal to the C=O surface dipole density on PET (0.235/nm$^2$).[35] From the fit, we find an adsorption energy of $\Delta G$ = -9 (±2) kJ/mol (see SI). This adsorption energy is indeed a few times smaller than the usual adsorption energy for molecular species adsorbed at an interface from solution.[36]

**General consideration for optimization and possible gate control on electricity generation devices.**

From what we learned about NaCl(aq.)/graphene/polymer interface, we can come up with the following general picture (sketched in Fig. 3e) for electricity generation by such devices. Ions in water are attracted to graphene by the surface dipole layer, if present, of a polymer and tend to adsorb in the interfacial dipole potential traps. As a drop of ionic solution moves on graphene, the fresh water/graphene interface is formed at the advancing water front. Sodium ions must rush over to the fresh interface. They increase the potential seen by electrons in graphene and attract them toward the fresh interface. Only in delayed action, the negative ions in the solution are dragged along by the positive ions and affect somewhat the current in graphene. This dynamic process constitutes the charging action. The opposite process occurs at the trailing edge of the drop and constitutes the discharging action. As proposed earlier,[5] electricity generation results from combined actions of charging and discharging. The above picture suggests that polymers with a denser and stronger surface dipole layer should be more efficient in generating electricity. A quick estimate (described in Section S5 in SI) shows that the generated current can be increased by ~10 folds if dipole moment is doubled and its surface density is four times larger than that of C=O in PET. Besides it is already known that larger drop size, faster drop velocity, and higher salt concentration generally enhance the current generation, while electricity generation also varies with the type of ions in solution as the dynamics of ion motion depends on the size and charges of the ions.[5]

It may happen that the surface dipole layer of a polymer is so strong that ions adsorbing on it (or on graphene of the graphene/polymer film) do not desorb at the receding edge. However, such a surface would be passivated by counter ions adsorbed from air. Accordingly, electricity generation efficiency of the device will be reduced. As a demonstration, we replace the PET substrate by a 1-mm z-cut $LiNbO_3$ ferroelectric crystal. The $LiNbO_3$ surface generates a much stronger surface field than PET, but exposing it to air rapidly decreases its surface potential. As a result, the voltage spikes generated by moving water drops on graphene/$LiNbO_3$ are slightly larger than in the PET case, (see Fig. S5 in the SI).

The $LiNbO_3$ case above suggests that the efficiency and durability of electricity generation of the device would be greatly improved with the use of ferroelectric films instead of surface-dipole films if surface passivation of ferroelectric films could be prevented. Because surface passivation comes from field-attracted ions, it is possible that they can be removed by switching off the ferroelectric polarization. This led us to the idea of constructing a switchable electricity generation device using a gate-controlled ferroelectric film. To demonstrate the idea, we used a thin film (40 nm) of $\beta$-polyvinylidene fluoride (PVDF, -(-$CF_2$-$CH_2$-)$_n$-) to replace the PET film in our device. $\beta$-PVDF is a well-known organic ferroelectric material with its domain consisting of ordered chains and the side groups (-$CF_2$-$CH_2$-).[37] The macroscopic ferroelectric polarization can be poled by an external electric field. In the device we studied (Fig. 4a), the $\beta$-PVDF film was sandwiched between graphene and 100nm silica grown on doped Si substrate. Graphene and Si were used as the gate electrodes for poling $\beta$-PVDF. The film was initially weakly polarized and was poled to become more strongly polarized by a gate voltage of 30V on the two electrodes. The two polarization states were monitored by SFVS showing a strong methylene stretch peak in the latter but a very weak one in the former, as seen in Fig.4b. The generated voltage spikes, depicted in Fig.4c, were strong and weak, respectively, in the two cases. Reversing the voltage reversed the poling of PVDF and switched the polarity of electricity generation. Thus, with appropriate gate voltages, we could switch electricity generation on and off. In the off state, ion adsorption at the interface is weak, and we should be able to clean the interface *in situ* by various means. Electricity generation of our device with PVDF was found to be 5 times stronger than the one with PET under the same working condition. Note that the device can be further improved through engineering of the PVDF film, for example, reduction of the switching voltage of the film.

Electricity generation devices using a polymer substrate with a surface dipole layer may have the disadvantage that their efficiency is limited by the short-range dipole field at the interface. Longer range surface field enabling attraction of more ions toward the interface to increase the efficiency is desirable with solutions of low salt concentration. This calls for preparation of a substrate with an appropriate surface charge layer in the device. Or alternatively, the longer surface field penetration can be achieved by patterning a ferroelectric substrate into sections with not too large lateral size to thickness ratio. As described in Fig. S6 of the SI, ferroelectric disks with radius of 100nm and thickness of 40nm has the field penetration length of 100nm, which is much longer than that for a surface dipole layer.

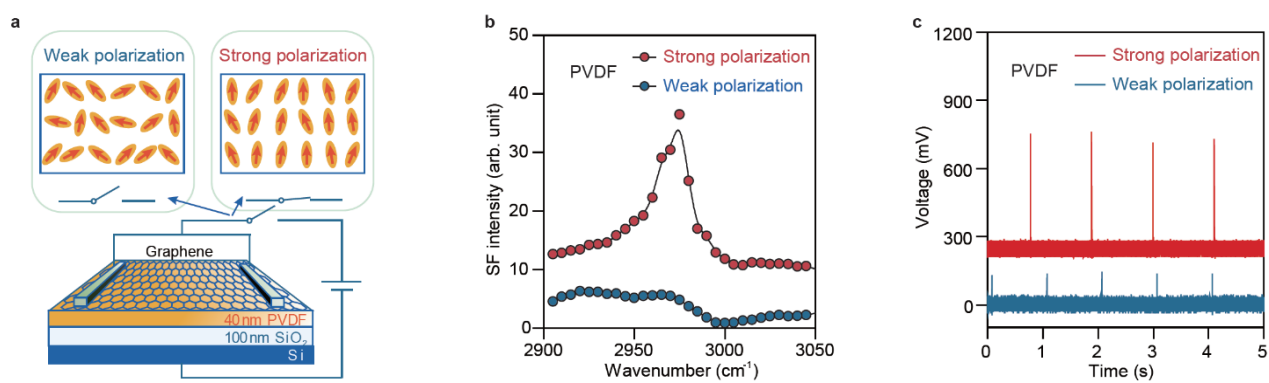

**Figure 4 Gate-control switchable voltage generation with a ferroelectric polymer film.** (a) Sketch of a device made of a graphene/PVDF/SiO$_2$/Si structure that can be gated across the layers by a bias voltage. The domain alignment of PVDF film along the surface normal can be controlled by the bias. (b) CH$_2$ stretching mode at 2975 cm$^{-1}$ monitored by SFVS, indicating that PVDF is in the weakly polarized state (blue dots) or in the stronger polarized state (red dots). (c) Measured voltage spikes from the device showing that they are weak when PVDF is in the weakly polarized state and strongly enhanced when PVDF is in the stronger polarized state.

In summary, graphene/polymer electricity generation devices are most promising because of their simplicity, flexibility, and scalability. Understanding of the underlying mechanism provides us with basic design concepts for possible improvement. Our finding that graphene does not attract ions and serves only as a conducting pathway for electricity generation suggests that it could be replaced by other conducting materials as long as their field-screening effect is weak. For better electricity generation efficiency of the solution/graphene/substrate devices, we suggest, for the substrate, use of polymers with stronger surface dipoles and dipole density, organic and inorganic ferroelectric films, and nano-patterned surfaces to improve ion attraction to the interface, as well as electrical gating to control and switch electricity generation.

## Acknowledgements

CST acknowledges support from the National Key Research and Development Program of China (No. 2016YFA0300902 and No. 2016YFC0202802), the National Natural Science Foundation of China Grants (No.11290161 and No.11374064) and Research Fund for the Doctoral Program of Higher Education of China (No. 20130071110023).

## Method

**Sample Preparation**

In constructing the graphene/PMMA/fused silica device, PMMA was first spin-coated on graphene CVD-grown on a copper foil. A fused silica window was then pressed on PMMA and annealed at 150 °C for 15min to have the window stick together with PMMA. The assembly was dipped into a 0.1M $(NH_4)_2S_2O_8$ solution for 2 hours to etch away the copper foil, followed by rinsing with deionized water, leaving the graphene surface uncontaminated. In the case of PET, a 10-micron PET film was heated to the glass phase, and a silica window was used to press it on graphene/copper foil. After cooling, the copper foil was etched away by the 0.1M $(NH_4)_2S_2O_8$ solution. For the PVDF device, the traditional graphene transfer procedure was used.[21]

The metal electrical contacts on graphene were secured by silver epoxy, and were covered by silica gel to prevent them from contacting the ionic solution.

**Sum frequency spectroscopy**

For SF intensity, $|\chi^{(2)}_{S,eff}(\omega)|^2$, measurement, the SFVS setup was similar to those described earlier.[38] A picosecond Nd:YAG (Ekspla) laser with 20 Hz repetition rate was used to generate a visible beam at 532 nm and a tunable IR beam between 2800 and 3800 cm$^{-1}$. The two beams overlapping on the sample had pulse energies and beam spot diameters of 50 μJ and 1.5 mm and 50 μJ and 1.0 mm, respectively. The SF output was normalized to that from a z-cut quartz.

For phase-sensitive SFVS measurement,[39] the same input beams were used, but they propagated collinearly through a reference y-cut quartz plate and onto the sample at an incident angle of 45°. The SF signal generated from the y-cut quartz interfered with that from the sample in the reflected direction and provided the phase information about the SF output. $\text{Im}\,\chi^{(2)}_{S,eff}(\omega)$ was then deduced from the measured $|\chi^{(2)}_{S,eff}(\omega)|^2$ and the phase of $\chi^{(2)}_{S,eff}(\omega)$.

Supplementary Information for

# Mechanism of Electric Power Generation from Ionic Droplet Motion on Polymer Supported Graphene

*Shanshan Yang[1,†], Yudan Su[1,†], Ying Xu[1], Qiong Wu[1], Yuanbo Zhang[1,2], Markus B. Raschke[3], Mengxin Ren[4], Yan Chen[5], Jianlu Wang[5], Wanlin Guo[6], Y. Ron Shen[1,7] and Chuanshan Tian[1,2,]\**

[1] Department of Physics, State Key Laboratory of Surface Physics and Key Laboratory of Micro- and Nano-Photonic Structures (MOE), Fudan University, Shanghai, 200433, China

[2] Collaborative Innovation Center of Advanced Microstructures, Nanjing, 210093, China

[3] Department of Physics, Department of Chemistry, and JILA, University of Colorado, Boulder, CO 80309, United States

[4] School of Physics and TEDA Applied Physics Institute, Nankai University, Tianjin, 300071, China

[5] National Laboratory for Infrared Physics, Shanghai Institute of Technical Physics, Chinese Academy of Science, Shanghai, 200083, China

[6] Institute of Nanoscience, Nanjing University of Aeronautics and Astronautics, 29 Yudao Street, Nanjing 210016, China.

[7] Department of Physics, University of California, Berkeley, CA 94720, United States

## S1. Preparation of fresh graphene surface on PMMA or PET

Our sample preparation procedure is depicted in Fig. S1.

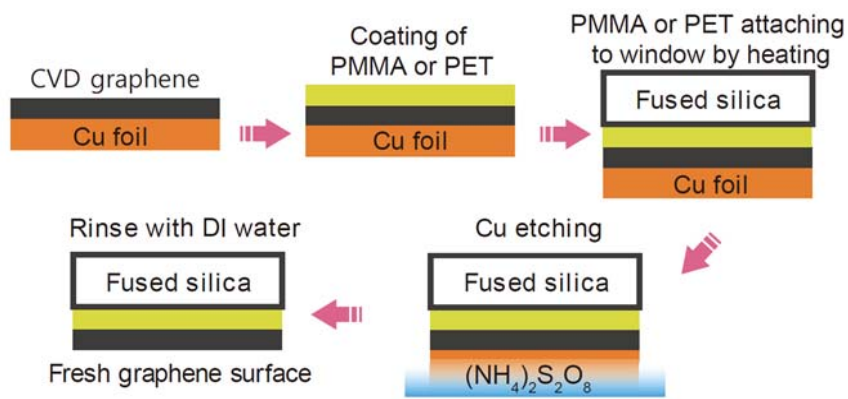

**Figure S1. Procedure for transferring graphene onto PMMA (PET) supported by a fused silica window.**

## S2. Raman spectra of graphene

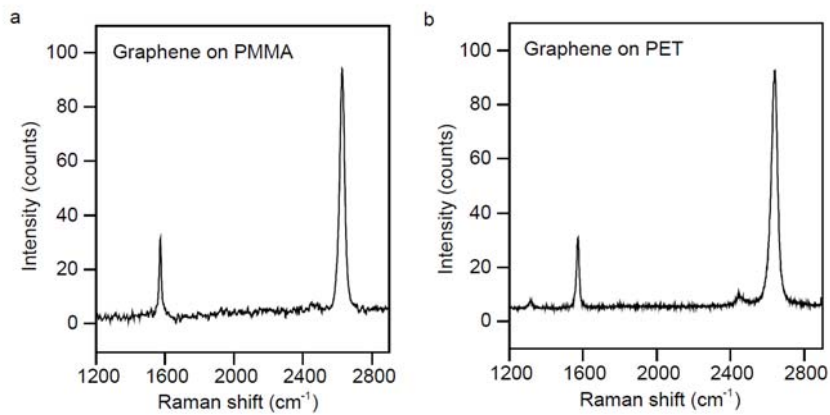

**Figure S2. Raman spectra of graphene on (a) PMMA and (b) PET**

The Raman spectra of CVD-grown graphene were measured under ambient conditions illuminated by a 632.8 nm HeNe Laser. The intensity ratio of the G mode at 1573 cm$^{-1}$ to the 2D mode at 2637 cm$^{-1}$ is 0.37 for graphene/PMMA and 0.35 for graphene/PET, indicating they were single layer graphene of good quality.[1]

## S3. Deduction of surface ion density from the complex $\chi^{(2)}$ spectrum

The effective surface nonlinearity for the charged water interface has the expression given by Eq.(1) in the main text and is reproduced below.

$$\chi^{(2)}_{S,\text{eff}}(\omega) = \chi^{(2)}_{S} + \int_{0^+}^{\infty} \chi^{(3)}_{B} E_{DC}(z) e^{i\Delta k_z z} dz \qquad (S1)$$

In our case, $\chi^{(2)}_{S}$ is for the interface without ion adsorption, but it is hardly changed when ion adsorption is not more than a few percent. The second term in the equation describes change of $\chi^{(2)}_{S,\text{eff}}$ induced by adsorbed ions, and comes from reorientation of water molecules by the surface field in the electric double layer (or diffuse layer). The complex $\chi^{(3)}_{B}$ is characteristic of bulk water and is known from previous measurement;[2] the phase mismatch $\Delta k_z$ is known from the beam geometry; and $E_{DC}(z) = -\nabla\phi(z)$ is related to the surface ion density σ by the Gouy-Chapman model in our analysis, with

$$\phi(z) = \frac{4k_B T}{e} \tanh^{-1}\left(\tanh\left(\frac{e\phi_0}{4k_B T}\right)\exp(-z/\lambda_D)\right) \qquad (S2)$$

$$\lambda_D = \left(\frac{\varepsilon_0 \varepsilon_r k_B T}{2ce^2}\right)^{1/2}$$

$$\phi_0 = -\frac{2k_B T}{e} \sinh^{-1}\left(\frac{e\sigma}{(8k_B T \varepsilon_0 \varepsilon_r c)^{1/2}}\right)$$

Here, $\lambda_D$ is the Debye length, $c$ and $\varepsilon_r$ are the ionic strength and the relative dielectric constant of the solvent, respectively. Using Eqs. (S1) and (S2) with σ as an adjustable parameter to fit the experimentally measured $\chi^{(2)}_{S,\text{eff}}(\omega) - \chi^{(2)}_{S}$ or its imaginary part allows us to deduce σ for given bulk NaCl concentration.

**S4. Deduction of adsorption free energy from Langmuir model.**

We used the simple Langmuir model to fit the experimental data of σ versus bulk NaCl concentration in water.[3]

$$\sigma = \sigma_{sat} \cdot \frac{K \cdot f}{1 + K \cdot f} \qquad (S3)$$

where $f$ is the molar fraction of Na$^+$ in water, and $K=\exp(-\frac{\Delta G}{N_A k_B T})$ with $\Delta G$ and $N_A$ being the adsorption free energy and the Avogadro number, respectively. Since Na$^+$ ions supposedly adsorb on top of the surface dipoles of PET, we assumed, in our fitting, $\sigma_{sat}$ equal to the surface dipole density, 0.235 e/nm$^2$,[4] and the fit displayed in Fig. 3e in the main text yielded $\Delta G$= -9±2 kJ/mol. When we took different values of $\sigma_{sat}$ ranging from 0.235 e/nm$^2$ to 0.062 e/nm$^2$ (obtained with 600 mM of NaCl in water), $\Delta G$ was found to vary from -9 to -12 kJ/mol, as seen from the fits in Fig. S3.

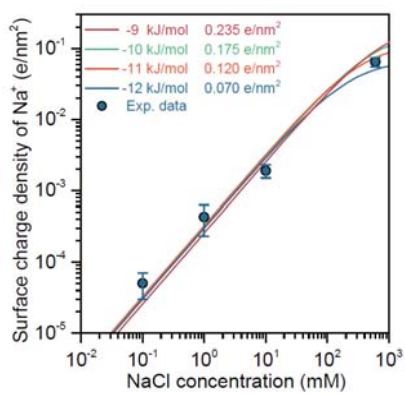

**Figure S3 Langmuir model fitting of surface charge densities of Na$^+$ ions versus NaCl concentrations**.

## S5. Electric potential from an ordered square C=O dipole array

To calculate the potential distribution near the water/PET interface generated by the surface dipoles of PET, we assumed an array of dipoles sitting on a square lattice with the following parameters (Fig. S4a): the normally oriented surface dipoles are characterized partial charge $\delta$=0.47$e$ and length $l$=1.2Å; the lattice constant is $a$=2nm; an interfacial layer between PET and water that has a permittivity varying linearly with distance from 2 of PET to 78 of water at water/PET interface, The calculated results are presented in Fig. 3b, S4b and S4c, showing highly localized potential wells around the dipoles.

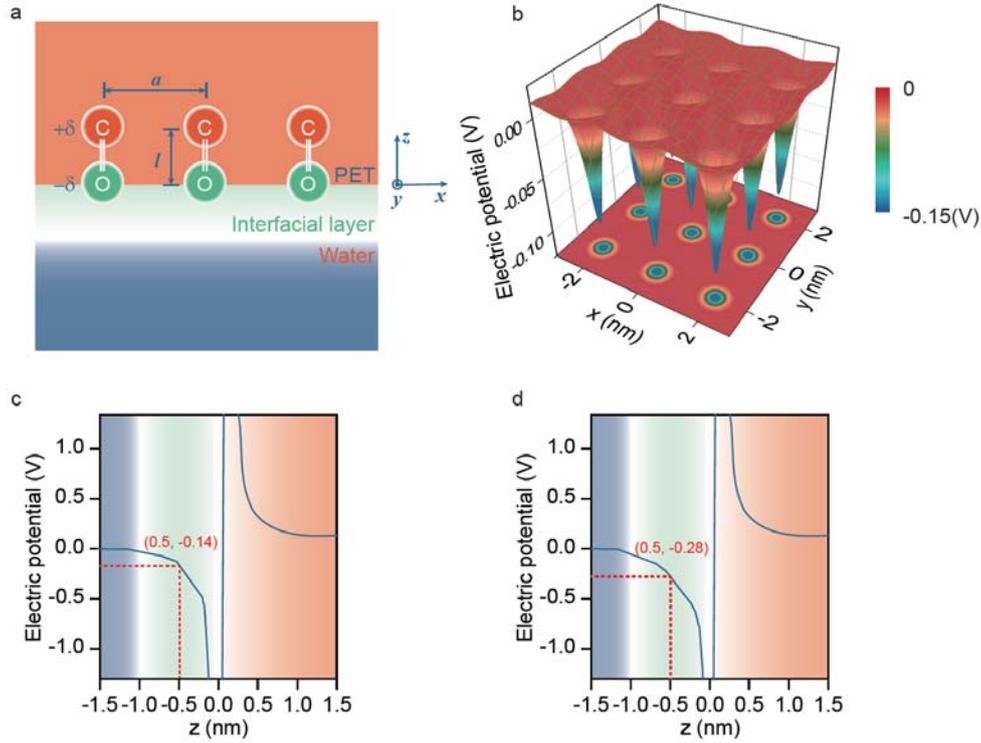

**Figure S4** (a) Model for COMSOL simulation of interfacial potential from a square array of dipoles at PET/water interface. (b) The electric potential distribution in the *xy* plane in water at 0.5 nm away from oxygen of the dipole layer. Electric potential (blue curve) along the surface normal (c) at PET/water interface and (d) ethylene carbonate/water interface. The red dashed lines in (c) and (d) mark the electric potentials at 0.5nm from the oxygen-atom layer on top of each dipole.

Knowing the dipole moment and surface density of C=O group on PET are moderate in terms of attracting ions from solution, we calculate in the following how much electric current can be improved if a dipole layer with stronger dipole and larger density is used in the device. The generated electric current is given by $I = \frac{\Delta Q}{\Delta t} \propto \frac{\sigma}{\Delta t}$,[5] where ΔQ is the change of charge in the pseudocapacitor formed at the interface, which is proportional to surface density (σ) of the adsorbed $Na^+$ ions. We consider a dipole layer consisting of ethylene carbonate with a molecular dipole moment of 4.9 Debye[6] and dipole density of $1/nm^2$. Assuming the normally oriented surface dipoles with partial charge $\delta$=1e and length $l$=1Å (4.9 Debye), the electric potential at z=0.5nm becomes -0.28V as shown in Fig. S4d, twice of that in PET case (Fig. S4c). Accordingly, the enthalpy change ΔU is -26 *kJ/mol* and -14 *kJ/mol* respectively in the two cases. For simplicity, we also assume the entropy change TΔS is the same in these two cases. Given the free energy for $Na^+$ ion adsorption at the water/graphene/PET interface being -9 *kJ/mol*, ΔG for $Na^+$ ion adsorption at a dense ethylene carbonate dipole layer is found to be ~-20 *kJ/mol*. Using Eq. S3, we estimate the surface density of adsorbed ions at room temperature is about 0.6 $e/nm^2$ under 600mM, which 10 folds of the PET case. Thus, the generated current from such device can be improved by one order of magnitude if the supporting substrate for graphene adopts the parameters of the surface dipole layer assumed above.

## S6. Electricity generation from graphene/LiNbO$_3$

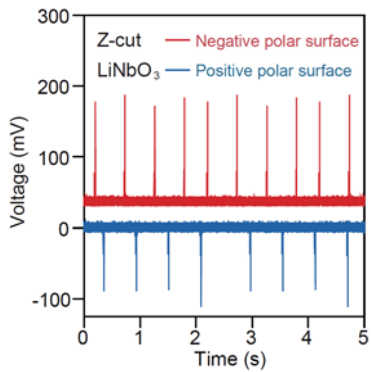

**Figure S5 The voltage traces generated from graphene/LiNbO$_3$ device.** The two curves are from graphene on the negative (red) and positive (blue) polar surface of LiNbO$_3$.

A 1 mm z-cut ferroelectric lithium niobate (LiNbO$_3$) wafer poled along the z-axis presumably can generate a much stronger field than PET. Our measurement with LiNbO$_3$ replacing PET in the device generated positive and negative voltage spikes from the opposite polar surfaces of LiNbO$_3$ as displayed in Fig. S5. They are only slightly stronger than those generated with PET, indicating that the field from LiNbO$_3$ was suppressed by counter ion adsorption on the surfaces.

## S7. Electric field distribution of a dipole disk

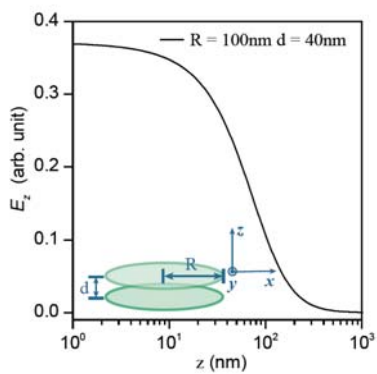

**Figure S6 Electric field distribution of a dipole disk.** Electric field distribution generated by a dipole disk with radius of 100nm along z above the disk center. Inset shows the schematic model for a dipole disk.

## S8. The screening effect of 3-5 layer graphene on PET

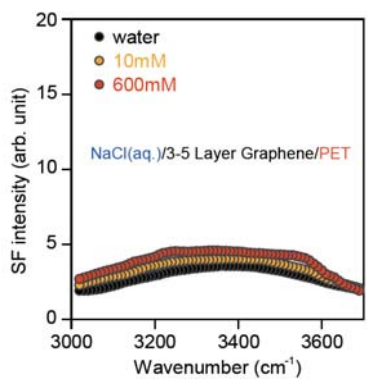

**Figure S7 SFVS spectra of NaCl(aq.)/3-5 layer graphene/PET interface.**

The screening of dipole field by multilayer graphene is expected to be larger than monolayer graphene. Indeed, as shown in Fig. S7, the SF intensity spectrum from 3-5 layer graphene/PET device increases a little when adding NaCl in water up to 600mM. The spectral enhancement in Fig. S7 is clearly weaker than that in Fig. 2d in the main text. It suggests that the dipole field is better screened by multilayer graphene that results in less adsorption of solvated ions at the interface than the monolayer case. As noted by Yin et. al.,[5] the electricity generation becomes less efficient if thicker graphene is used. For quantitative comparison, we plot in Fig. 3c enhancement of the SF intensity at 3300 cm$^{-1}$ of NaCl(aq.)/graphene/PET interface with respect to that of pure water for monolayer and 3-5 layer of graphene, respectively.